\documentclass[conf]{new-aiaa}
\usepackage[utf8]{inputenc}

\usepackage{graphicx}
\usepackage{amsmath}
\usepackage{MnSymbol}
\usepackage[version=4]{mhchem}
\usepackage{siunitx}
\usepackage{longtable,tabularx}
\usepackage{color}
\usepackage{algorithm}
\usepackage{algpseudocode}

\usepackage{caption}
\usepackage{subcaption}
\usepackage[colorlinks=true, linkcolor=black, citecolor = black, urlcolor = black,filecolor=black , pagebackref=false,hypertexnames=false, plainpages=false, pdfpagelabels]{hyperref}
\usepackage[numbers,sort&compress]{natbib}  
\allowdisplaybreaks

\newcommand{\bm}[1]{\mathbf{#1}}

\graphicspath{{./}{figures/}}

\title{Partial Replanning for Decentralized Dynamic Task Allocation}

\author{Noam Buckman \footnote{S.M. Candidate, Dept. of Mechanical Engineering, MIT, Cambridge, MA, nbuckman@mit.edu} and Jonathan P.~How\footnote{Richard Cockburn Maclaurin Professor of Aeronautics and Astronautics, MIT, Cambridge, MA, jhow@mit.edu}}
\affil{Laboratory for Information and Decision Systems \\ Massachusetts Institute of Technology, Cambridge, MA 02139}

\author{Han-Lim Choi\footnote{Assistant Professor, Div. of Aerospace Engineering, KAIST, Daejeon, Korea, hanlimc@kaist.ac.kr}}
\affil{Korea Advanced Institute of Science and Technology, Daejeon, Korea}

\begin{document}

\maketitle
\begin{abstract}
	In time-sensitive and dynamic missions, multi-UAV teams must respond quickly to new information and objectives.  This paper presents a dynamic decentralized task allocation algorithm for allocating new tasks that appear online during the solving of the task allocation problem. Our algorithm extends the Consensus-Based Bundle Algorithm (CBBA), a decentralized task allocation algorithm, allowing for the fast allocation of new tasks without a full reallocation of existing tasks.  CBBA with Partial Replanning (CBBA-PR) enables the team to trade-off between convergence time and increased coordination by resetting a portion of their previous allocation at every round of bidding on tasks.  By resetting the last tasks allocated by each agent, we are able to ensure the convergence of the team to a conflict-free solution.  CBBA-PR can be further improved by reducing the team size involved in the replanning, further reducing the communication burden of the team and runtime of CBBA-PR.  Finally, we validate the faster convergence and improved solution quality of CBBA-PR in multi-UAV simulations.
\end{abstract}

\section*{Nomenclature}

{\renewcommand\arraystretch{1.0}
	\noindent\begin{longtable*}{@{}l @{\quad=\quad} l@{}}
		$n_r$  & number of robots \\
		$n_t$ &   number of tasks \\		
		$ \mathcal{I}$ & set of robots \\
		$ \mathcal{J}$ & set of tasks \\
		$L_t$& Maximum length of path \\
		$D$ & network diameter \\
		$T^*$ & new task \\
		$\bm{b}_{i}$& bundle \\
		$\bm{p}_i$ & path \\
		$y_{ij}$ & winning bids \\
		$z_{ij}$ & winning agents \\
		$\mathcal{J}_{reset}$ & subset tasks reset in replan\\
		$(i^*,j^*)$ & optimal assignment in central greedy solution \\
\end{longtable*}}

\section{Introduction}

The use of UAVs and UAGs in large teams has become increasingly desired and viable as robot hardware has decreased in size and cost.  Likewise, there is increasing interest in solving large, more complex missions that require multi-agent teams to accomplish a varied number of tasks.  Decentralized algorithms have allowed planners to scale with larger team sizes, amortizing computation and communication across the robot teams.  In addition, decentralized algorithms, which only rely only peer-to-peer communication, can be used in environments without a communication infrastructure or in environment with constrained centralized communication.  For example, a team of UAVs operating in a foreign terrain, may not have access to classic communication infrastructure that one may be accustomed to in local settings, especially for missions utilizing airspace or underwater environments.  Likewise, in an adversarial setting, where opponents may look to target a central planner, decentralized algorithms provide robustness to single-point failures caused by a central planner or communication infrastructure. 

This article investigates the decentralized dynamic task allocation problem where a team of robots must respond to new tasks that appear during the mission, allocating a new task with its existing allocations.  This is in contrast to the static task allocation problem which assumes that all the tasks are known before the team executes the task allocation solver.   The problem is similar to other NP-hard problems such as the Dynamic Vehicle Routing Problem (D-VRP) or Dial-A-Ride Problem \cite{Psaraftis1980}, where online requests occur during the operation of the vehicles, in which new locations must be visited by the vehicles.  In addition, we specifically seek a decentralized algorithm that relies only on peer-to-peer communication to ensure robustness and scalability. 

In a centralized setting, such as those studied in the operations research and logistics communities, solvers have been developed to provide heuristics for searching the space of solutions in the dynamic vehicle routing problem.  Ref.~\cite{Pillac2013} and \cite{Ritzinger2016} provide excellent reviews on dynamic VRP solutions.  The first group of approaches is to periodically replan, rerunning the static task allocation solver at predetermined time epochs, such as in the ant colony algorithm \cite{Montemanni2005}.  The second group of approaches is to continuously generate plans to create a shared pool of possible solution, from which a solution can be adapted when a new customer arrives.  These algorithms include the adaptive memory algorithm  \cite{Ichoua2007a} and genetic algorithms \cite{VanHemert2004}, however, they rely on a centralized memory or global situational awareness.  In \cite{Edison2008,Guangtong2018}, the genetic algorithm is extended to multiple UAVs, however they fail to be fully decentralized as a central planner is still required.

As for fully decentralized algorithms, most have focused on convex optimization or task-assignment where the task score functions are independent. Ref.~\cite{Nedic2009} successfully decentralizes the cooperative optimization by reaching consensus on sub-gradients, however, optimizes a convex score function with continuous decision variables.  Ref.~\cite{Chopra2017} introduced a decentralized version of the Hungarian algorithm for task assignment, however, requires that the task scores are independent.  Ref.~\cite{Liu2013a} presents an online solver by enforcing strict task swapping, but again relies on the task assignment problem where scores are independent. As for a decentralized planner for the combinatorial optimization in task allocation, \cite{Choi2009} introduces the CBBA algorithm which can provide an approximate solution to the vehicle routing problem when all the tasks are introduced at the beginning of the algorithm.  This article extends the work in \cite{Choi2009} to adapt to new tasks while maintaining solution quality and convergence.

In this work, we propose CBBA with Partial Replanning (CBBA-PR) which quickly allocates the new task by only reallocating a subset of tasks.  Where as the static decentralized solver Consensus-Based Bundle Algorithm (CBBA) requires a full re-solving of the original task allocation problem to allocate a new task, CBBA-PR allows for a partial replanning of the existing allocations.  This is achieved by enabling agent to partially reset their allocation between rounds of auctioning.  We show that this partial resetting strategy still converges to a conflict-free solution.  In addition, the amount of resetting can be chosen to achieve a desired response time for the system.  In doing so, the team has the flexibility to allow for little coordination but quick response, or vice versa.  We also present a method for choosing a subset of robots to participate in the reallocation process.  Finally, we validate the convergence of CBBA-PR and solution quality improvements, compared to the baseline CBBA approach.  

The remainder of this paper is structures as follows.  In Section II, we state the dynamic task allocation problem statement and describe the Consensus-Based Bundle Algorithm, which we build off of in this paper.  In Section III, we describe and analyze CBBA's existing approaches to allocating a new task.  Section IV, presents the main algorithm: CBBA with Partial Replanning, a resetting approach that guarantees quicker allocation of the new task.  Section V reports simulations results that show improvements in convergence and solution quality.  Finally, in Section VI we provide concluding thoughts and future directions.

\section{Decentralized Task Allocation: Consensus-Based Bundle Algorithm (CBBA)}
\subsection{Problem Statement}
The goal of the static task allocation problem is to allocate a set of $n_{t}$ tasks to $n_{r}$ agents to arrive at a conflict-free assignment of tasks to robots.  Generally, the agents can be assigned up to $L_t$ tasks which can represent either a physical limitation or a planning horizon for the agent. The decentralized task assignment can then be formed as an optimization:
\begin{equation*}
\begin{aligned}
& \max  & &  \sum_{i=1}^{n_{r}} \left(  \sum_{j=1}^{n_{t}} c_{ij}(\bm{x}_i,\bm{p}_{i})x_{ij} \right) \\
& \text{subject to:}  & & \sum_{j=1}^{n_{t}} x_{ij} \leq L_t \quad \forall i \in  \mathcal{I} \\	
& 							  & &\sum_{i=1}^{n_r} x_{ij} \leq 1 \quad \forall j \in  \mathcal{J} \\
& 							  & & \sum_{i=1}^{n_r} \sum_{j=1}^{n_t} x_{ij}  = \min \{n_rL_t,n_t\} \\
& 							  & & x_{ij} \in \{0,1\}, \quad \forall(i,j) \in \mathcal{I} \times \mathcal{J} \\
\end{aligned}
\end{equation*}
where $x_{ij}=1$ if agent $i$ is assigned to task $j$ and $\mathbf{x_i}$ is a vector of length $n_t$ with the assignment of all tasks in $\mathcal{J}$.  The variable length vector $\bm{p_i}$ represent the path for agent $i$ which is a list of the tasks assigned to agent $i$ in order of execution.  The current length of the path is $|\bm{p}_i|$ and is not allowed to be longer than the path constraint $L_t$.  

In the dynamic scenario, a new task $T^*$ arrives during or at the end of the task allocation process.  Now the agents must allocate a total of $n_t+1$ tasks.  We denote the new set of tasks $\mathcal{J}'$, new paths $\bm{p}_1' \dots \bm{p}_i'$, and new decision variables, $x_{ij}'$.  The team must now optimize the following optimization:

\begin{equation*}
\begin{aligned}
& \max  & &  \sum_{i=1}^{n_{r}} \left(  \sum_{j=1}^{n_{t}+1} c'_{ij}(\bm{x}_i',\bm{p}_i')x'_{ij} \right) \\
& \text{subject to:}  & & \sum_{j=1}^{n_{t}+1} x'_{ij} \leq L_t \quad \forall i \in  \mathcal{I} \\	
& 							  & &\sum_{i=1}^{n_r} x'_{ij} \leq 1 \quad \forall j \in  \mathcal{J'} \\
& 							  & & \sum_{i=1}^{n_r} \sum_{j=1}^{n_t} x'_{ij}  = \min \{n_rL_t,n_t+1\} \\
& 							  & & x'_{ij} \in \{0,1\}, \quad \forall(i,j) \in \mathcal{I} \times \mathcal{J}' \\
\end{aligned}
\end{equation*}

\subsection{Consensus-Based Bundle Algorithm (CBBA)}
Consensus-Based Bundle Algorithm \cite{Choi2009} is a decentralized auction based algorithm designed to solve the static task allocation problem, where all the task are known at the beginning.  The algorithm alternates between two main phases: the \textit{bundle building} phase and the \textit{consensus} phase of the algorithm.  In the bundle building phase, the agents iteratively generate a list of tasks to service by bidding on the marginal increase for each task.  In the consensus phase, the agents resolve differences in their understanding of the winners of each task.  Before proceeding, we define five lists used in the running of CBBA:
\begin{enumerate}
	\item A \textit{path}, $\bm{p}_i \triangleq \{ p_{i1},\dots p_{i|\bm{p}_i|} \}$ is a list of tasks allocated to agent $i$.  The path is in the order by which agent $i$ will service the tasks.
	\item A corresponding \textit{bundle}, $\bm{b}_{i} \triangleq \{ b_{i1},\dots b_{i|\bm{b}_{i}|} \}$ is the list of tasks allocated to agent $i$ in the order by which agent $i$ \textit{bid} on each task, i.e. task $b_{im}$ is added before $b_{in}$ if $m < n$ .  The size of $\bm{b}_{i}$, denoted $|\bm{b}_{i}|$ cannot exceed the size of $\bm{p}_i$ and an empty bundle is denoted $\bm{b}_{i} = \emptyset$.
	\item A list of winning agents $\bm{z}_i \triangleq \{ z_{i1} \dots z_{in_t} \} $, where each element $z_{ij} \in \mathcal{I}$ indicates who agent $i$ believes is the winner of task $j$ for all tasks in $\mathcal{J}$.  If agent $i$ believes that no one is the winner of task $j$, then $z_{ij} = -1$.
	\item A corresponding list of winning bids $\bm{y}_i \triangleq \{ y_{i1} \dots y_{in_t} \}$ where $y_{ij}$ is agent $i$'s belief of the highest bid on task $j$ by winner $z_{ij}$ for all $j$ in $\mathcal{J}$.  If agent $i$ believes that no one is the winner of task $j$, then $y_{ij} = -\infty$.
	\item A list of timestamps $\bm{s}_i \triangleq \{ s_{i1}, \dots s_{i n_{r}} \} $ where each element $s_{ik}$ represents the timestamp of the last information that agent $i$ received about a neighboring agent $k$, either directly or indirectly.
\end{enumerate}

\subsubsection{Phase 1:  Bundle Building}
Unlike other algorithm which enumerate every possible allocation of tasks for agent $i$, in CBBA the agents greedily bid on a bundle of tasks.  In the bundle building phase (Algorithm \ref{alg:BundleBuild}), an agent $i$ determines the task $J_i$ that will yield the maximum increase in marginal score when inserted into its previous path.  If this score is larger than the current team winner, agent $i$ will add the task $J_i$ to its bundle. This process is repeated until it can no longer add tasks to its path, concluding by updating its list of winners and bids, $\bm{z}_{i}$ and $\bm{y}_i$.

\begin{algorithm}[t]
	\caption{CBBA Phase 1: Bundle Build}
	\label{alg:BundleBuild}
	
	\begin{algorithmic}[1]
		\State $ \bm{y}_i(t) = \bm{y}_i(t-1) $
		\State $ \bm{z}_i(t) = \bm{z}_i(t-1) $	
		\State $ \bm{b}_{i}(t) = \bm{b}_{i}(t-1) $ \label{lst:line:bundleReset}		
		\State $ \bm{p}_i(t) = \bm{p}_i(t-1) $ \label{lst:line:pathReset}	
		
		\While{$ |\bm{b}_{i}(t)| < L_t $}
		\State $ c_{ij} = \max_{n \leq |\bm{p}_i(t)|+1}  S_i^{\bm{p}_i(t) \oplus_n {j}}-S_i^{\bm{p}_i(t)}, \forall j \in \mathcal{J} \setminus \bm{b}_{i}(t)$
		\State $h_{ij} = I(c_{ij} \geq y_{ij}), \forall j \in \mathcal{J}$
		\State $J_i = \arg \max_j c_{ij}\cdot h_{ij}$
		\State $n_{i,J_i} = \arg \max_n  S_i^{\bm{p}_i(t) \oplus_n {J_i}}$
		\State $\bm{b}_{i}(t) = \bm{b}_{i}(t) \oplus_{end} {J_i}$
		\State $\bm{p}_i(t) = \bm{p}_i(t) \oplus_{n_i,J_i} {J_i}$
		\State $y_{i,J_i}(t) = c_{i_,J_i}$
		\State $z_{i,J_i}(t) = i$
		\EndWhile
		
	\end{algorithmic}
	
\end{algorithm}

\subsubsection{Phase 2:  Consensus}
In the second phase of CBBA, each agent $i$ communicates their updated lists, $\bm{z}_i, \bm{y}_i$ and $\bm{s}_i$ to their neighboring agents and resolve any conflicts in their belief of winners.  An important aspect of this process is that if two neighbors disagree on a specific task $\bar{j}$ located at location $\bar{n}_i$ in their bundles, the two agents are required to reset not only task $\bar{j}$ but also any tasks located in the bundle after $\bar{n}_i$:
\begin{equation} 
\begin{aligned}
& y_{i,b_{in}}=-\infty , \qquad z_{i,b_{in}}=-1 \quad \forall n > \bar{n}_i \\
& b_{in} = \emptyset, \quad n \geq \bar{n}_i \\ 
\end{aligned}
\end{equation}
where $b_{in}$ denotes the $n$th entry of bundle $\bm{b}_{i}$ and $\bar{n}_i = \min \{ n: z_{i,b_{in}} \neq i \}$.  The resetting of subsequent tasks is necessary for the proper convergence of CBBA, as the bids for those subsequent tasks ($y_{i,b_{in}}$) were made assuming a bundle consisting of the reset task $\bar{j}$.

\subsection{Convergence of CBBA}
Along with providing a procedure for decentralized allocation, Choi et. al. were able to show that CBBA converges in $O(n_t D)$ rounds of communication, where $D$ is the network diameter, and that CBBA arrives at the same result as a centralized sequential greedy algorithm (SGA).  In addition, they showed that for submodular value function, the sequential greedy solution achieves 50\% of the optimal score.   To prove convergence and optimality of the algorithm, CBBA requires that the score function has diminishing marginal gains (DMG). This leads to decreasing scores within an agent's own bundle ($y_{b_{in},j} \geq y_{b_{im},j} \: \forall n > m$), a characteristic of the bidding that also leads to CBBA's convergence. They show in Lemmas 1 and 2 \cite{Choi2009} that during the running of CBBA the team sequentially agree on the SGA solution.  Specifically, after $O(nD)$ rounds of communication, the team will agree on the first $n$ tasks allocated using a sequential greedy allocation ($j_1^*, j_2^*, \dots j_n^*$).  Also, the bids for the task will be optimal, $y_{i,j^{*}_{n}} = c^{*}_{ij^{*}_{n}} \: \forall i \in \mathcal{I} $, and the agents will remain in agreement on those scores for the duration of the task allocation.

\section{Bundle Resetting in Consensus-Based Bundle Algorithm}
The Consensus-Based Bundle Algorithm was originally intended for the static task allocation, in that it guarantees convergence when the tasks are known initially.  The authors \cite{Johnson2011} proposed that in dynamic settings, when information is outdated or there is a large change in situational awareness, the team should re-solve the new task allocation problem by rerunning CBBA. The shortcoming of this approach, however, is that in missions with a large number of tasks $n_t$ and a network diameter $D$, the response time for a new task will be $O(n_tD)$ for a single task.  In addition, a full re-solving of CBBA ignores the fact that the team had already arrived at a conflict-free solution, wasting the computation and communication used to allocate the original $n_t$ tasks.

For a quick response, one could allow for absolutely no replanning, without allowing any resetting of an agent's previous allocation, $\bm{p}_i(t-1)$, $\bm{b}_i(t-1)$.  This approach, which we will call CBBA with \textit{No Bundle Reset}, was in the original version of CBBA \cite{Choi2009}, having the Bundle Build process begin each round with $\bm{p}_i(t)=\bm{p}_i(t-1)$ and $\bm{b}_i(t) = \bm{b}_i(t-1)$. The advantage of CBBA with No Bundle Reset is that the convergence of the algorithm is virtually unaffected by the new task.  For example, in the case where the team has already reached convergence on the original $n_t$ tasks and arrived at some allocations $\bm{p}_1, \dots, \bm{p}_i$, the agents will never consider reallocating their existing tasks and simply bid on inserting the new task into their existing bundles $\bm{p_i'} = \bm{p}_i \oplus T^*$.  By effectively only bidding on $T^*$ and not allowing any bidding on other tasks in its paths, the team is able to reach agreement very quickly in $O(D)$ time.  While it is beyond this paper to provide quality guarantees for the no reset solution, intuitively it is clear that a no reset solution provides very little flexibility to the robot team in allocating $T^*$.  For example, in a highly constrained systems where many robots are at capacity $|\bm{p}_i(t-1)|=L_t$ or there are only a few robots that can service specific tasks, then only those robots under capacity and with the ability to service $T^*$ will be considered for $T^*$.  In these constrained scenarios, robot teams will need reset their previous allocations to consider the new task.  

A later addition to CBBA was to begin the Bundle Build process by fully resetting the previous allocations, $\bm{b}_i(t) \rightarrow \emptyset $ and $\bm{p}_i(t) \rightarrow \emptyset$ \cite{Johnson2012}.  This approach, CBBA with \textit{Full Bundle Reset}, gives the agents maximum flexibility in allocating the new task, in that they are not bound by their previous allocations.  While this full bundle reset increases the team coordination, one possible shortcoming of any bundle resetting approach is that it will no longer guarantee convergence for the original task allocation problem, as the algorithm is introducing additional resetting at each round of Bundle Build.

\textit{Claim:} If all tasks are known at the beginning of CBBA, both CBBA with Full Bundle Reset and CBBA with No Bundle Reset arrive at the SGA solution in $O(n_tD)$

\textbf{Proof:}
CBBA's convergence to the centralized sequential greedy algorithm's (SGA) solution relies on the fact that at some time $t$ the team will agree on the first $n$ tasks in the SGA solution and then subsequently agree on this solution for the rest of time (Lemma 1 \cite{Choi2009}).  The authors use induction to show that the team will first agree on the highest valued task (the first task allocated in the greedy solution) and after $nD$ rounds of communication, will agree on the first $n$ tasks in the SGA solution (Lemma 2  \cite{Choi2009}). In the case of a full reset at the beginning of Bundle Build, we need to show that the reset will not break Lemma 1, i.e. that if the team agrees on the first $n$ SGA tasks, they will continue to agree on those tasks for $s>t$.  First, denote the list of agreed SGA tasks at time $t$, as $\mathcal{J}^*_{(n)} = j^{*}_{(1)} \dots j^*_{(n)}$ and the SGA winners of those tasks as $i^*_{(1)} \dots i^*_{(n)}$.  Note that according to Lemma 1, at time $t$, all agents are in agreement on the bids fo the first $n$-SGA tasks: 
\begin{equation}
y_{ij} = c^*_{ij} \quad \forall j \in \mathcal{J}_* \quad \forall i \in \mathcal{I}
\end{equation}
As such, at some time $t$, agent $i$ will have a bundle $b_i$ that consists of agreed-on SGA tasks $\bm{b}_{i}[:n^*_i](t)$, where $n^*_i$ is the number of tasks in $\mathcal{J}^*_{(n)}$ that are assigned to agent $i$ by the SGA solution.  The rest of the bundle will consist of other tasks from $\mathcal{J}$ that may or may not be in consensus with the rest of the team, $\bm{b}_{i}[n^*_i+1:](t)$.  At time $t+1$, when the agent resets its bundle at the beginning of Bundle Build, it will begin greedily choosing tasks from $\mathcal{J}$ to add to its now empty bundle.  However, when agent $i$ calculates its own bid on a task $j_{(k)}^*$ in $\mathcal{J}^*_{(n)}$ where $i^*_{(k)} \neq  i$ (i.e. for tasks whose SGA winner is not $i$), agent $i$ will always be outbid the current team winner since their bids are greedily optimal. Instead, agent $i$ will first re-assign itself any of the tasks in $\mathcal{J}^*_{(n)}$ that have $i$ as the SGA winner, since those tasks will have the highest bids for agent $i$ by definition, since they are the centralized sequential greedy bids.  As a result, after the full bundle reset the agent $i$ rebuilds its first $n^*_i$ in its previous bundle, $\bm{b}_{i}(t+1)[:n^*_i] = \bm{b}_{i}(t)[:n^*_i]$.  This means that even in a full bundle reset, Lemma 1 and Lemma 2 hold, and thus convergence to the SGA is guaranteed in $O(n_tD)$.

We have just shown that a full reset and no reset converge to the same solution, however, when a new task is introduced, these two approaches diverge in terms of solutions and convergence guarantees.  First, in the proof above, the full reset converged to the sequential greedy solution because the Bundle Build process rebuilds the first part its previous bundle $\bm{b}_{i}(t)[:n^*_i]$, even after fully resetting its allocation.  However, if a new task is now considered in the building process, agent $i$ is not guaranteed to rebuild $\bm{b}_{i}(t)[:n^*_i]$.  In fact, it may be the case that the sequential greedy solution for $n_t+1$ tasks, $\mathcal{J'}^*$, will be completely different to the solution for original static $n_t$ task allocation problem.  Thus a full reset may result in a completely new allocation, requiring a full $O(n_tD)$ rounds of communication, even for a single new task.  In summary, CBBA's existing approaches to allocating a new tasks is either to to allow a full rerunning of CBBA (full reset), requiring $O(n_tD)$ rounds of communication, or a quick consensus on a winner for the new task, without allowing any reallocation of the existing tasks (no reset).

\section{CBBA with Partial Replanning (CBBA-PR)}
\subsection{Partial Resetting of Local Bundles}

\begin{figure}[t]
	\begin{subfigure}[t]{0.25\linewidth}
		\includegraphics[width=.99\columnwidth]{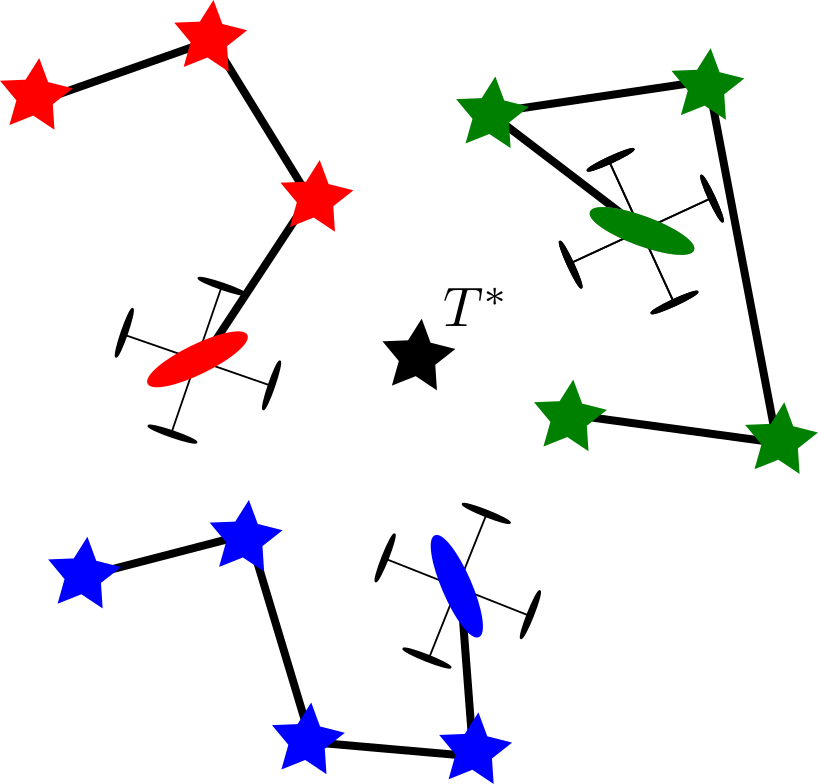}
		\caption{ Initial bundles $\bm{b}_1 \dots \bm{b}_i$ and new task $T^*$}
		
	\end{subfigure}	\hspace{0.1\linewidth}%
	\begin{subfigure}[t]{0.25\linewidth}
		\includegraphics[width=.99\columnwidth]{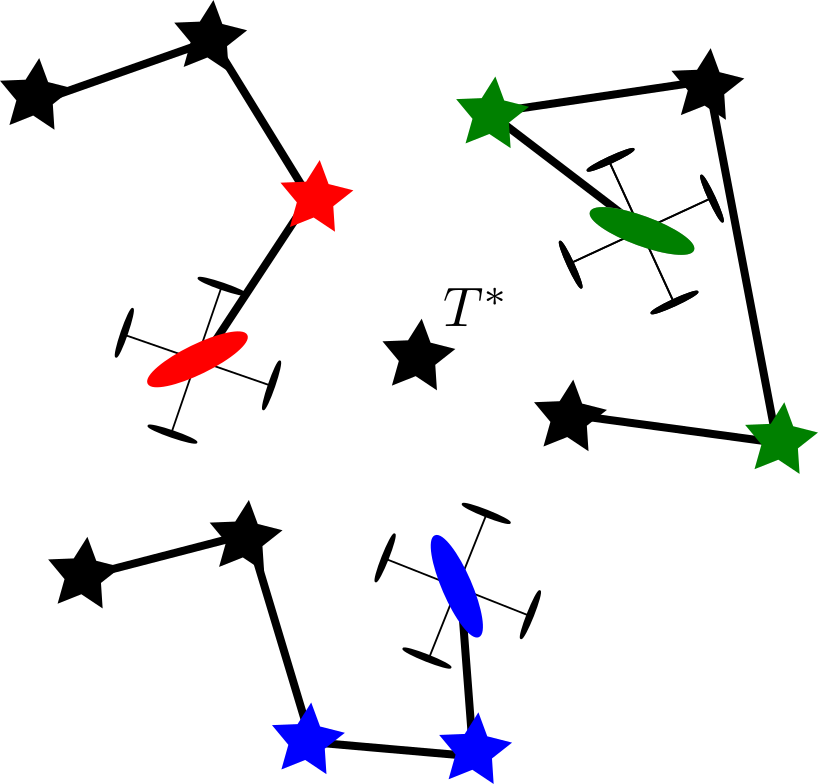}
		\caption{Each agent resets lowest $n_{i,reset}$ tasks in bundle}
	\end{subfigure}\hspace{.1\linewidth}%
	\begin{subfigure}[t]{0.25\linewidth}
		\includegraphics[width=.99\columnwidth]{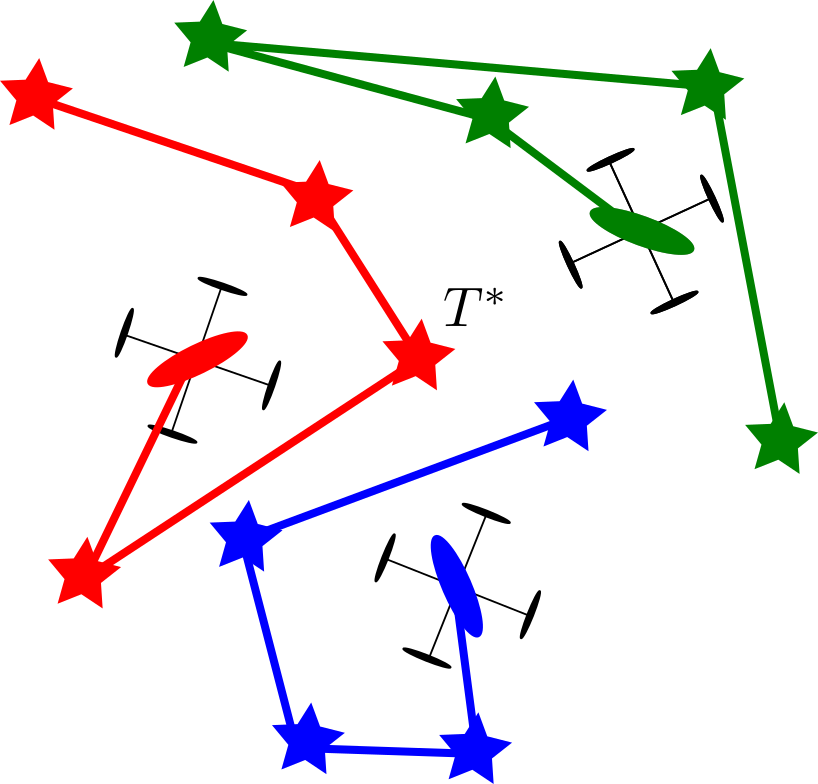}
		\caption{Converges to modified allocations with $T^*$}
	\end{subfigure}
	\label{fig:replanDrawing}
	\caption{Dynamic task allocation using CBBA-PR by partially resetting the last task in each agent's bundle at the beginning of Bundle Build.  The tasks are chosen to be the last tasks auctioned in the bundle (not the order of physical path) to ensure convergence of CBBA-PR}
\end{figure}
To better trade-off coordination with the speed of convergence, we propose CBBA with Partial Replan (CBBA-PR) which enable each agent to reallocate a portion of their existing allocation at each round of CBBA.  In CBBA-PR, each agent resets part of their bundle at the beginning of Bundle Build, releasing their $n_{i,reset}$ lowest bid tasks from their previous bundles.  The $n_{i,reset}$ can be chosen by the team depending on the amount of replanning or response speed that is necessary for the team.  For example, in the case where new tasks are frequently appearing and the team wants to converge before another new task arrives, they may choose $n_{i,reset}$ to be very small.  On the other hand, if the new tasks are particularly high-valued, the team can tallow for more coordination by selecting a larger number of tasks to reset.  Furthermore, the amount of resetting may change during the duration of CBBA.  If the new task arrives early on in the team's allocation of the original $n_t$ tasks, they may allow for more resetting.  While if the team has already converged on all $n_t$ original tasks, they may limit the amount of resetting, to not waste the computation for the original tasks.

An important requirement for the tasks chosen for resetting is they must be the \textit{lowest} tasks in each agent's respetice bundles.  This is to ensure the convergence of CBBA, for if tasks are reset in any other order (randomly chosen or maximum bids), CBBA will not have diminishing valued bids, and the team will not converge to a conflict-free solution.  Rather, if the agents reset only the lowest $n_{reset}$ tasks in each bundle to reset, we can re-use Lemmas 1 and 2 to prove that the team sequentially agree on a conflict-free solution.

\subsection{Improving on the Convergence of CBBA-PR}
One limitation of the local partial reset strategy is that while average convergence will generally be better than a full reset, we can not guarantee that worst-case performance will improve.  For example, if an agent only has one task to reset, and that task happens to be the first task in the centralized SGA solution, a full replan may occur.  However, if the team has converged on the first $n_t$ tasks before $T^*$ arrives, then we can guarantee worst-case performance of $O(n_{reset}D)$ where $n_{reset}=n_r \times n_{i,reset}$ is the \textit{total} number of tasks reset by the team. In this scenario, the team can choose the $n_{reset}$ lowest bid tasks from across the entire team.  Since the team has already reach consensus on the original centralized greedy solution, those $n_{reset}$ lowest solutions will in fact be the last $n_{reset}$ tasks allocated by the SGA. Since the higher bid tasks will remain allocated after the partial reset, the team is guaranteed to converge within $O(n_{reset}D)$ rounds of communication,

In this procedure, CBBA with Partial Team Replan (Algorithm \ref{alg:PartialTeam}), when a new task appears, each agent sorts the final bid array $\bm{y}_i$, enabling the agents to identify the $n_{reset}$-lowest SGA tasks, $\mathcal{J}_{reset}$ (Line \ref{alg:partialTeam:sort}).  Any agent with a task from $\mathcal{J}_{reset}$ in their previous bundle, will reset the task by removing it from $\bm{b}_i$ and $\bm{p}_i$ and resetting the values in $\bm{y}_i$ and $\bm{z}_{i}$.  By doing so, the team is able to get increased coordination from reallocating existing tasks while still guaranteeing convergence that is $O(n_{reset}D)$, where $n_{reset}$ can be chosen to fit the team's desired response time.  In addition, if only a subset of the team $\mathcal{I}_{reset}$ is chosen to participate in the replanning, the team can reuse the known assignments in $\bm{z}_{i}$ to specifically reset $n_{reset}$ tasks that were assigned to agents in $\mathcal{I}_{reset}$, ensuring that none of the reset tasks are ``wasted" on agents that are not participating in the replan.  Conversely, the team can choose a combination of $n_{reset}$ tasks and desired subteam of diameter $d$, reusing $\bm{y}_i$ and $\bm{z}_i$ to achieve replanning within a desired convergence.  With this subteam and subtask selection, the team can choose between selecting a large subteam with few tasks per robot to reallocate or a small subteam with robots fully resetting previous allocations.  In general, this ideal mix of $d$ and $n_{reset}$ for a given scenario will be dependent on the mission characteristics.

\begin{algorithm}[t]
	\caption{CBBA-PR with Partial Local Replan (Fixed Bundle Size)}
	\label{alg:PartialLocal}
	
	\begin{algorithmic}[1]
		\State $ \mathcal{J}_{i,reset} = \{  b_{im}(t-1) \: \forall m \geq n_{reset} \} $
		\ForAll{$j \in \mathcal{J}_{i,reset}$} 
			\State $\bm{b}_i(t) = \bm{b}_i(t-1) \ominus j$
			\State $\bm{p}_i(t) = \bm{p}_i(t-1) \ominus j$
			\State $z_{i,j}(t) = -1$
			\State $y_{i,j}(t) = \infty$
		\EndFor
		\State Phase 1: Bundle Build($\bm{p}_i(t),\bm{b}_{i}(t),\bm{y}_i(t), \bm{z}_i(t))$
		\State Phase 2: Consensus
	\end{algorithmic}
	
\end{algorithm}

\begin{algorithm}[t]
	\caption{CBBA with Partial Team Replan}
	
	\begin{algorithmic}[1]
		\State Given:  $\mathcal{I}_{reset}$, $t_{response}$
		\State $d = Diameter(\mathcal{I}_{reset})$ \label{alg:partialTeam:subteam}
		\State $n_{reset} = \frac{t_{response}}{d \times \Delta_{comm} }$ \label{lst:line:nReset}		
		\State $\bm{y}_i^s = Sort(\bm{y}_i)$ \label{alg:partialTeam:sort}
		\State $\mathcal{J}_{reset} = \bm{y}_i^s[n_{reset}:n_t]$
		\ForAll{$j \in \mathcal{J}_{reset}$}
		\State $ \bm{p}_i(t) = \bm{p}_i(t-1) \ominus j $	
		\State $ \bm{b}_{i}(t) = \bm{b}_{i}(t-1) \ominus j $
		\State $ y_{ij}(t) = - \infty$
		\State $ z_{ij}(t) = -1 $
		\EndFor
		
		\State Phase 1: Bundle Build($\bm{p}_i(t),\bm{b}_{i}(t),\bm{y}_i(t), \bm{z}_i(t))$
		\State Phase 2: Consensus
	
	\end{algorithmic}
	\label{alg:PartialTeam}
\end{algorithm}

\section{Results}

\subsection{Simulation}
\begin{figure}[t]
	\centering
	\begin{subfigure}{.3\linewidth}
		\includegraphics[width=0.99\columnwidth]{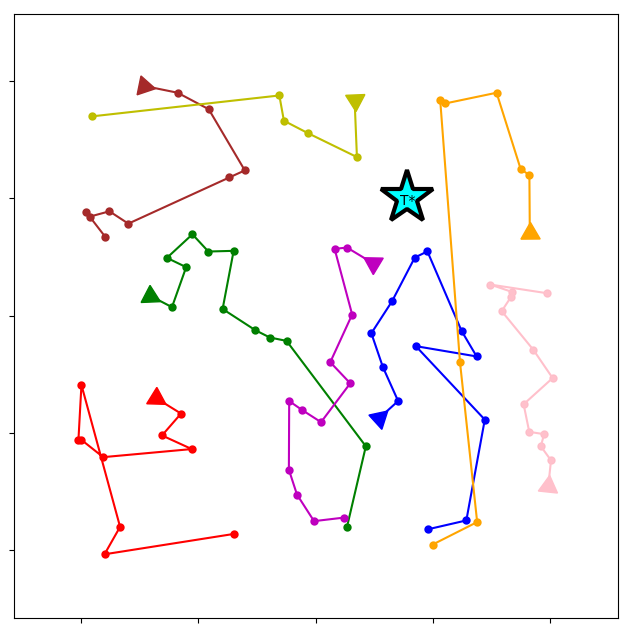}
	\end{subfigure}%
	\begin{subfigure}{.3\linewidth}
		\includegraphics[width=0.99\columnwidth]{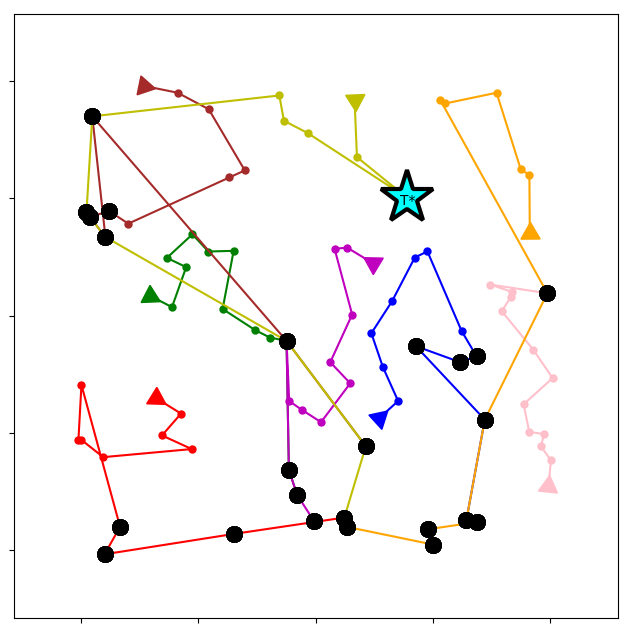}
	\end{subfigure}%
	\begin{subfigure}{.3\linewidth}
		\includegraphics[width=0.99\columnwidth]{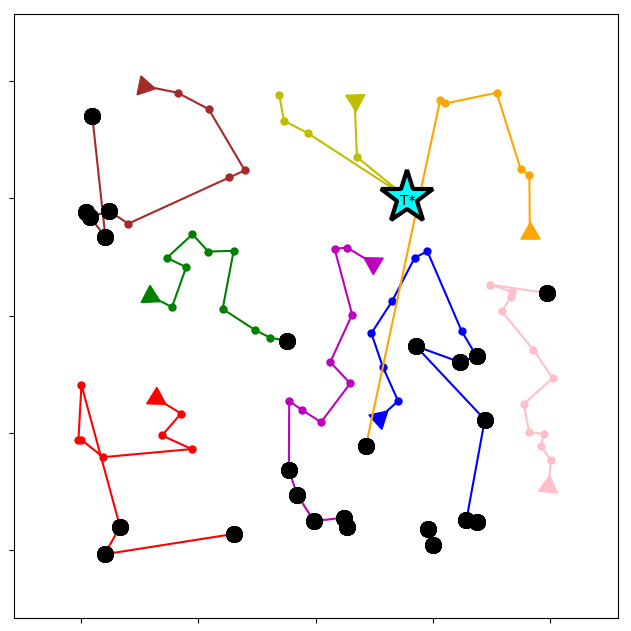}
	\end{subfigure}
	
	\begin{subfigure}{.3\linewidth}
		\includegraphics[width=0.99\columnwidth]{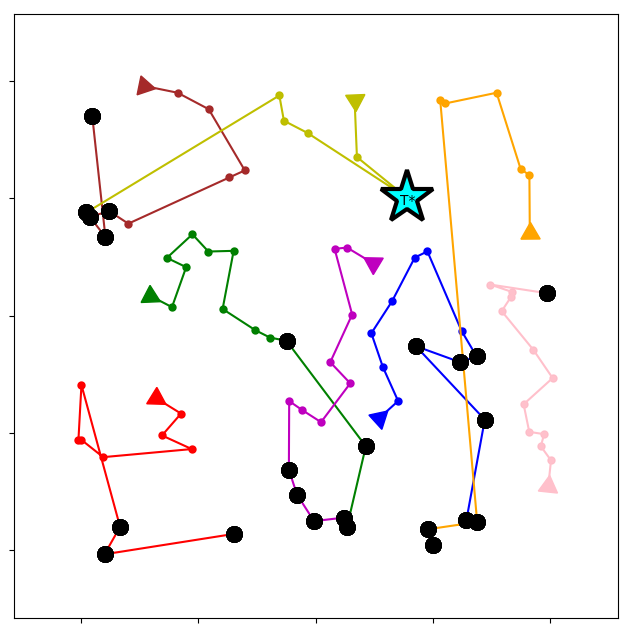}
	\end{subfigure}%
	\begin{subfigure}{.3\linewidth}
		\includegraphics[width=0.99\columnwidth]{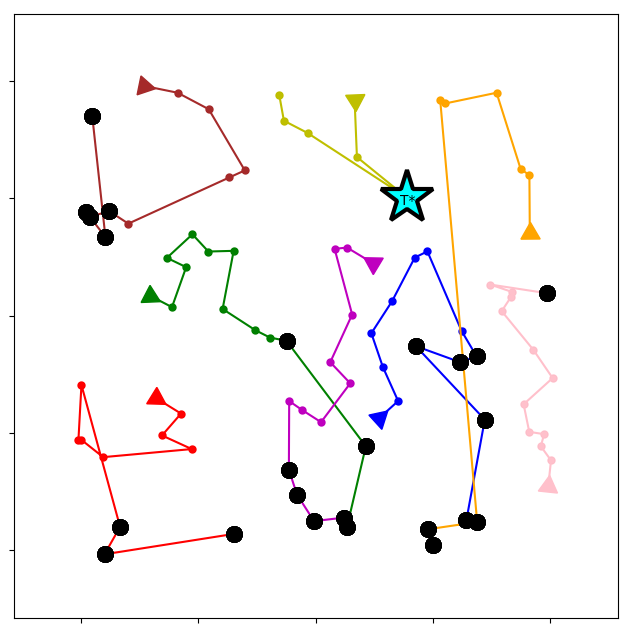}
	\end{subfigure}%
	\begin{subfigure}{.3\linewidth}
		\includegraphics[width=0.99\columnwidth]{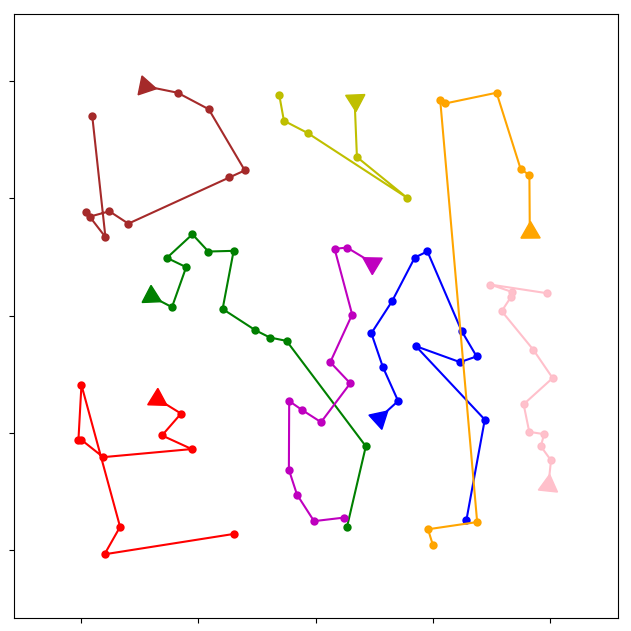}
	\end{subfigure}%
	\caption{Simulation of eight robots allocation $n_t=80$ tasks, allocated tasks $\bm{p}_i$ are colored corresponding to the assigned robot.  A new task $T^*$ (green star) appears sequentially and tasks are released (black, filled circles) until all are allocated.}
	\label{fig:SimExample}
\end{figure}

A UAV task allocation simulator was created to validate the convergence and quality of solutions for various replanning strategies.  The simulator is implemented in Python and allows for varying communication conditions, dynamic robot movements, and newly appearing tasks.  CBBA with Partial Replan is run locally on multiple instances of the Robot class and the Simulator only facilitates message passing between agents and the revealing of new tasks the team.  We implement a vehicle routing scenario where $n_r=8$ UAVs must visit $n_t=80$ task locations.  In these experiments, we use a time-discounted scoring function:
\begin{equation}
S(\bm{p}_i) = \sum \limits_{j \in \bm{p}_i} \lambda_{ij}^{\tau_j^{\bm{p}_i}} R_{ij}
\end{equation}
where $\lambda_{ij} \in (0,1]$ is the time-discount value, $R_{ij}$ is the static reward of task $j$ by agent $i$, and $\tau_j^{\bm{p}_{i}}$ is the time it takes to service task $j$ along path $\bm{p}_i$. We run 100 monte carlo simulations where the initial tasks are placed in randomly located location, initialized with $R_{ij}=1$ and $\lambda_{ij}=0.95$.  Once the team converges on an initial solution $\bm{p}_1 \dots \bm{p}_{i}$, a new task $T^*$ arrives that must be allocated by the team.  This process is repeated 8 times for a total arrival of 8 tasks.  For each simulation scenario, the setting is saved so that multiple strategies can be run and compared.  Figure \ref{fig:SimExample} shows an example simulation, where initially a new task appears (top left), then tasks are reset, and a final allocation is reached (bottom right). Note that significant changes and disagreement during the replanning phase since the team is resetting a subset of previous tasks while allocating the new task.

\subsection{Comparing Convergence}

 We compare the number of rounds of CBBA required for the team to agree on a conflict-free solution, using the four strategies outlined above: no bundle resetting, partial local bundle reset, partial team reset, and a full bundle reset.  In both cases of partial resetting, the team initially resets a total of $n_{reset}=24$ tasks, where the difference lies in resetting a fixed number from each bundle (local reset) or choosing the lowest tasks from the entire teem (team reset).
 We first compare the team's convergence for the initial static allocation of $n_t$ tasks in Figure \ref{fig:Convergence} (left) and then in Figure \ref{fig:Convergence} (right), the final team convergence time after a new task $T^*$ appears .  
 In the static allocation, all four strategies perform with equal convergence times as expected by the theory.
 When a new task is introduced and needs to be allocated by the team, all four strategies require increased rounds of CBBA, ranging from no reset with the least bidding to a full reset which requires the most rounds of CBBA.  Between the local and team resetting, the local performs worse, in some simulations, requiring the same number of rounds as a full reset.
 This is expected as only the worst case can be guaranteed when the lowest team wide tasks are chosen for resetting.  However, on average, the local bundle reset does perform faster than a full reset, suggesting that there is still a speed up from a partial local bundle reset.

\subsection{Comparing Solution Quality}
\begin{figure}[p]
	\centering
	\includegraphics[width=0.49\linewidth]{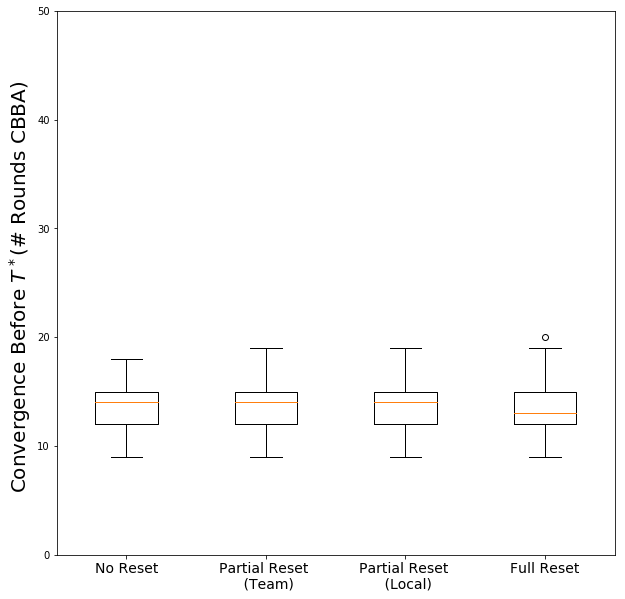}
	\includegraphics[width=0.49\linewidth]{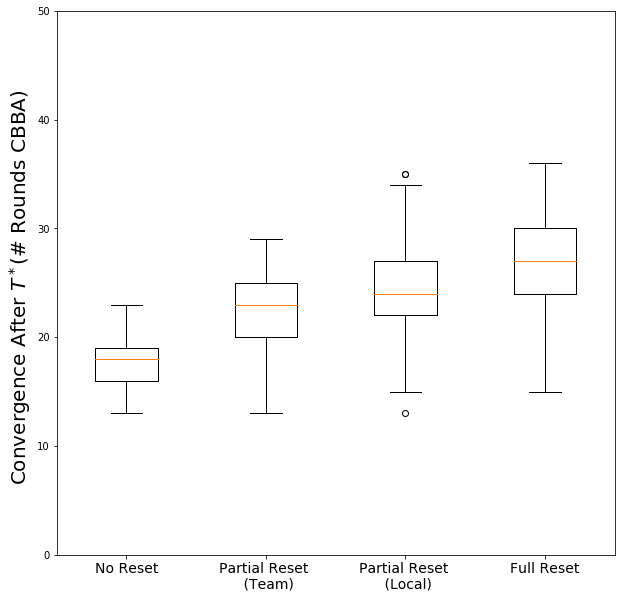}
	\caption{Convergence time for the initial static allocation (left), before the new task $T^* $ arrives, is the same for all four replan strategies.  When a new tasks arrives (right), the number of rounds on average and worst-case is highest for a full reset replan and shortest for the no reset strategy. Choosing the lowest-$n$ tasks to reset for a global replan converges faster than a fixed number of tasks reset in each bundle and provides intermediate performance as a whole.}
	\label{fig:Convergence}

	\centering
	\includegraphics[width=0.7\linewidth]{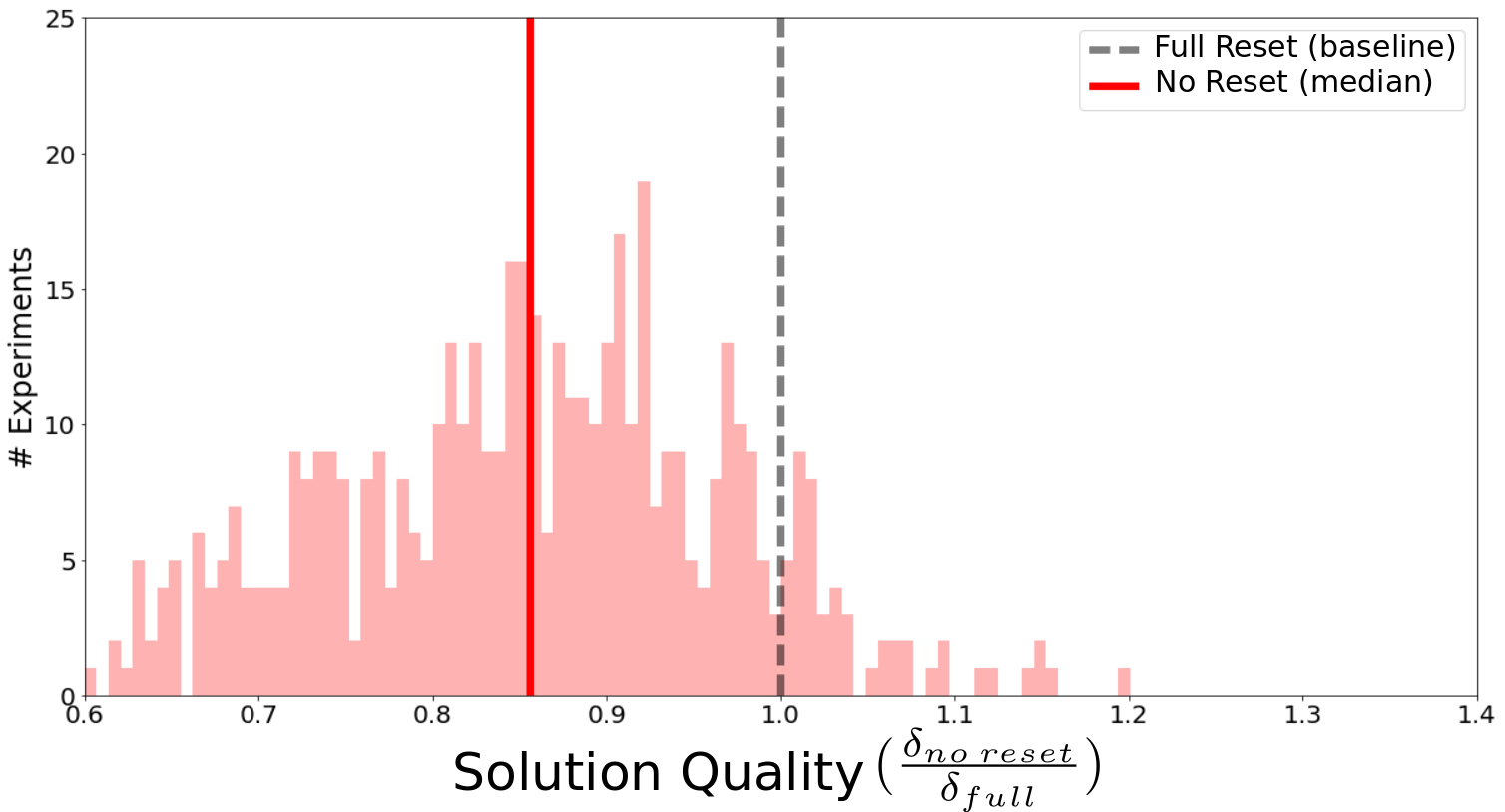}

	\centering
	\includegraphics[width=0.7\linewidth]{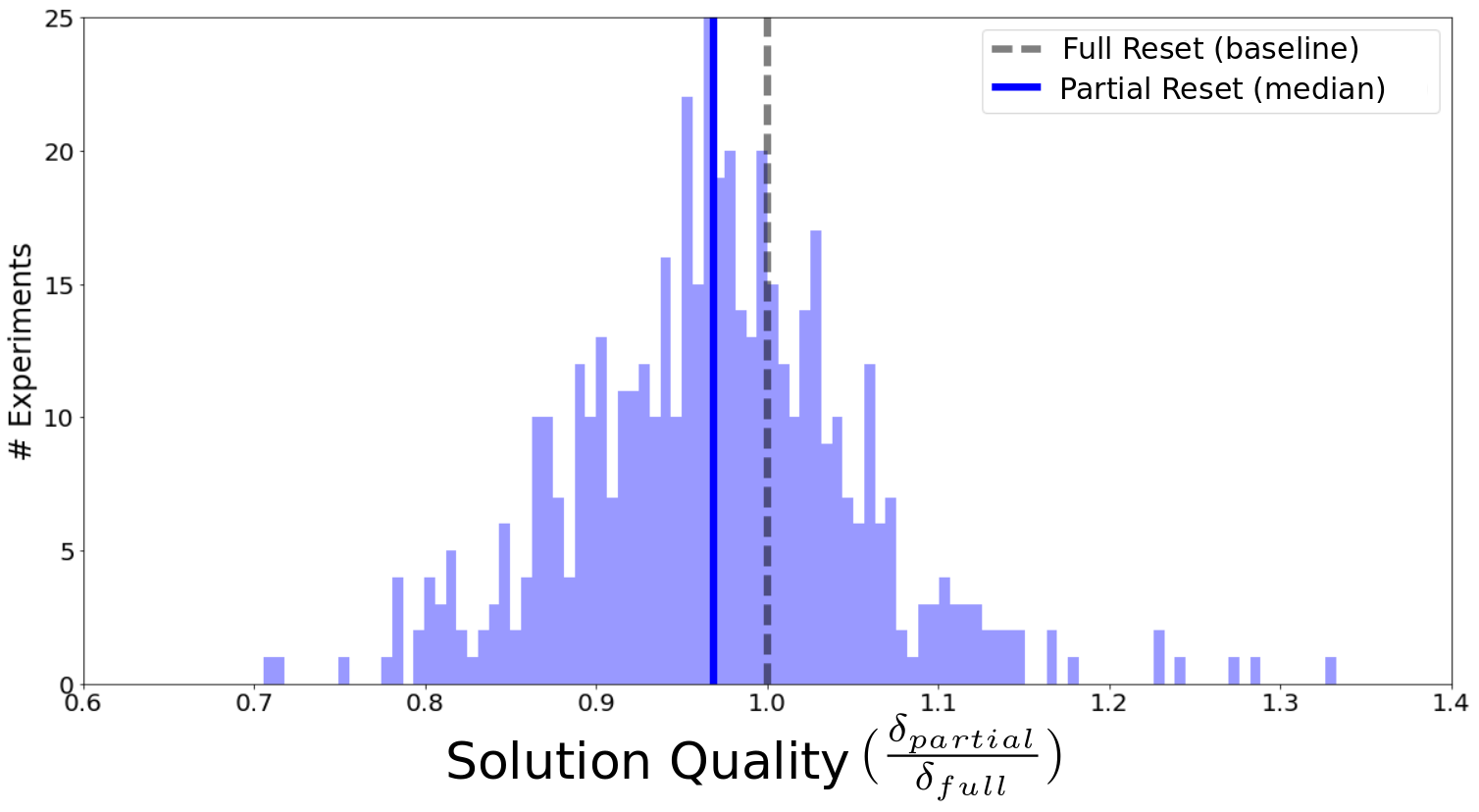}
	\caption{Performance of partial replanning compared to no replanning, measured by score increase after allocating 8 new tasks.  Partial replan improves the score quality, nearing the performance of full replan baseline.}
	\label{fig:SolutionQuality}
\end{figure}

To understand the performance gains of partial replanning, we compare the replan strategies to the full reset strategy.  While the full reset is not an optimal solution, we will use it as a baseline for "best" performance since it does have the 50\% approximation of CBBA and intuitively has the highest level of coordination.  In doing so, we compare the convergence of CBBA-PR compared to the full reset CBBA. The performance of each algorithm is measured by the increase on team score $\delta = \sum_{i \in \mathcal{J}} S_i(\bm{p}_i') - \sum_{i \in \mathcal{J}} S_i(\bm{p}_i')$ caused by servicing the new task, where $\bm{p}_i'$ is the solution \textit{after} all the new tasks are allocated. Figure \ref{fig:SolutionQuality} shows the performance of both no reset (top) and partial reset (bottom) in an unconstrained setting, i.e. $L_t n_r > n_t$.  As expected, the no reset and partial reset perform worst than the baseline full reset, however, the faster partial reset algorithm outperforms no resetting and generally performs more similar to a full reset.  Note that the high variance in solution quality is due to the full reset still being suboptimal due to its greedy nature.  However, in more constrained setting where the number of feasible solutions is fewer, partial and full reset will more consistently outperform no reset approaches.

\section{Conclusion}
In this work, we presented a dynamic task allocation algorithm that trades off the team's response time for solution quality.  By resetting the lowest bid tasks from previous rounds of CBBA, the team is able to get fast convergence while still coordinating with other agents.  In addition, if all original tasks are already allocated, the team can faster guaranteed convergence by selecting the team-wide lowest bid tasks, reducing the tasks allocated and number of agents involved.  Finally, simulations showed that the team could in fact get faster convergence than re-solving the task allocation problem and better solutions than no coordination.  This framework, trading off the time to re-solve the problem with new information, can be explored for other areas of optimization and planning.  In addition, future work may include responding to other levels of dynamics in the environment, such as the addition and loss of robots, outdated information, and time-varying task information.

\section*{Acknowledgments}
This work was supported by the Department of Defense (DoD) through the National Defense Science \& Engineering Graduate Fellowship (NDSEG) Program, and Lockheed Martin.  Thanks to Dr. Golnaz Habibi for the valuable insights.

\bibliography{main.bib}

\end{document}